\documentclass[]{aa} 
\usepackage{graphicx}
\usepackage{natbib}

\newcommand{\Teff}{\hbox{$T\sb{\rm eff}$}}          
\newcommand{\logg}{\hbox{$\log g$}}
\newcommand{\Msun}{\hbox{M$\sb{\odot}$}}

\begin{document}

   \title{The accretion-diffusion scenario for metals in cool white dwarfs}

   \author{D. Koester
	   \and 
           D. Wilken
           }
   \institute{   Institut f\"ur Theoretische Physik und Astrophysik, 
             University of Kiel, {D-24098 Kiel}, Germany
}

   \offprints{D. Koester\\ \email{koester@astrophysik.uni-kiel.de}}

   \date{}

\authorrunning{D. Koester \& D. Wilken}

\titlerunning{Metals in cool white dwarfs}

\abstract{We calculated diffusion timescales for Ca, Mg, Fe in hydrogen
     atmosphere white dwarfs with temperatures between 5000 and
     25000~K. With these timescales we determined accretion rates for a
     sample of 38 DAZ white dwarfs from the recent studies of
     \cite{Zuckerman.Koester.ea03} and
     \cite{Koester.Rollenhagen.ea05}. Assuming that the accretion
     rates can be calculated with the Bondi-Hoyle formula for
     hydrodynamic accretion, we obtained estimates for the interstellar
     matter density around the accreting objects. These densities are
     in good agreement with new data about the warm, partially ionized
     phase of the ISM in the solar neighborhood.
      }

\maketitle

\section{Introduction}
Gravitational separation of elements in the strong gravitational field
in the outer layers of white dwarfs has been known for a long time to
be the primary physical process determining the atmospheric
composition \citep{Schatzman47}. In the absence of competing processes
(stellar wind, radiative levitation, convection), the heavy elements
diffuse downward on a timescale that is always short compared to the
evolutionary timescale, leaving the lightest element that is present floating
on top. In most cases this is hydrogen, leading to the spectral type
DA; if no hydrogen is present, the atmosphere is pure He (spectral
types DB, DC).

\begin{table}[ht]
\caption{Data for 38 objects with detected photospheric
  Ca. Atmospheric parameters \Teff, \logg, and logarithmic Ca/H ratio,
  [Ca/H], are from \cite{Zuckerman.Koester.ea03},
  \cite{Koester.Rollenhagen.ea05}, or
  \cite{Berger.Koester.ea05}. Space velocities, $v$, relative to the
  sun are from \cite{Pauli03}, \cite{Zuckerman.Koester.ea03}. In eight
  cases where the space velocity is not known, we have replaced it
  with $\sqrt{3}$ times the radial velocity, taken from the cited
  reference. Column d is the distance in pc, taken from
  \cite{Pauli03}, \cite{McCook.Sion99}.
  \label{tab1}}
\begin{center}
\begin{tt}     
\begin{tabular}{lrrrrr}
Object&  \Teff & \logg & [Ca/H] & $v$       & d    \\ 
      &  [K]       &       &        &  [km/s]   & [pc] \\
\hline 
WD0032-175  &  9235 &  8.0 & -10.5  &  74  & 32 \\ 
HS0047+1903 &  16600&  7.8 & -6.1   &   5  &    \\
HE0106-3253 &  15700&  8.0 & -5.8   &  20  & 66 \\ 
WD0208+296  &   7201&  7.9 & -8.8   &  89  &    \\ 
WD0235+064  &  11420&  7.9 & -9.2   &  11  & 34 \\
WD0243-026  &   6798&  8.2 & -9.8   &  52  & 21 \\ 
WD0245+541  &   5190&  8.2 & -11.7  &  38  & 10 \\ 
HS0307+0746 &  10200&  8.1 & -7.1   &  38  &    \\
WD0408-041  &  14400&  7.8 & -6.6   &  40  & 74 \\ 
WD0543+579  &   8142&  8.0 & -10.5  & 120  & 30 \\
WD0846+346  &   7373&  8.0 & -9.4   &  20  & 27 \\
WD1015+161  &  19300&  7.9 & -5.9   &  47  & 95 \\ 
WD1102-183  &   8026&  8.0 & -10.3  &  45  & 40 \\ 
WD1116+026  &  12200&  7.9 & -6.5   &  26  & 43 \\ 
WD1124-293  &   9700&  8.1 & -8.2   &  65  & 34 \\ 
WD1150-153  &  12800&  7.8 & -6.0   &  14  & 85 \\ 
WD1202-232  &   8800&  8.2 & -9.8   &  37  & 10 \\ 
WD1204-136  &  11200&  8.0 & -7.2   &  56  & 52 \\ 
WD1208+576  &   5830&  7.9 & -10.8  &  65  & 20 \\ 
HE1225+0038 &   9400&  8.1 & -9.7   &  38  & 29 \\ 
WD1257+278  &   8481&  7.9 & -8.1   &  50  & 35 \\ 
HE1315-1105 &   9400&  8.4 & -9.4   &  17  & 40 \\ 
WD1337+705  &  20435&  7.9 & -6.7   &  53  & 33 \\ 
WD1344+106  &   6945&  8.0 & -11.3  &  79  & 18 \\ 
WD1407+425  &   9856&  8.0 & -9.8   &  21  & 26 \\
WD1455+298  &   7366&  7.6 & -9.3   & 101  & 35 \\ 
WD1457-086  &  20400&  8.0 & -6.1   &  24  & 117\\ 
WD1614+160  &  17400&  7.8 & -7.2   &  44  & 117\\ 
WD1633+433  &   6569&  8.1 & -8.6   &  30  & 15 \\ 
WD1821-131  &   7029&  8.4 & -10.9  &  89  & 20 \\ 
WD1826-045  &   9200&  8.1 & -8.6   &  26  & 25 \\ 
WD1858+393  &   9470&  8.0 & -7.8   &   3  & 35 \\
WD2115-560  &   9700&  8.1 & -7.6   &  54  & 22 \\
HS2132+0941 &  13200&  7.7 & -7.1   &  37  & 79 \\ 
WD2149+021  &  17300&  7.9 & -7.7   &   4  & 25 \\
HE2221-1630 &  10100&  8.2 & -7.2   &  25  & 52 \\ 
HS2229+2335 &  18600&  7.9 & -5.9   &  35  & 91 \\ 
WD2326+049  &  12100&  7.9 & -6.4   &  59  & 19 \\
\hline  
\end{tabular}
\end{tt} 
\end{center}
\end{table}

\begin{figure*}
\includegraphics[angle=270,width=18cm]{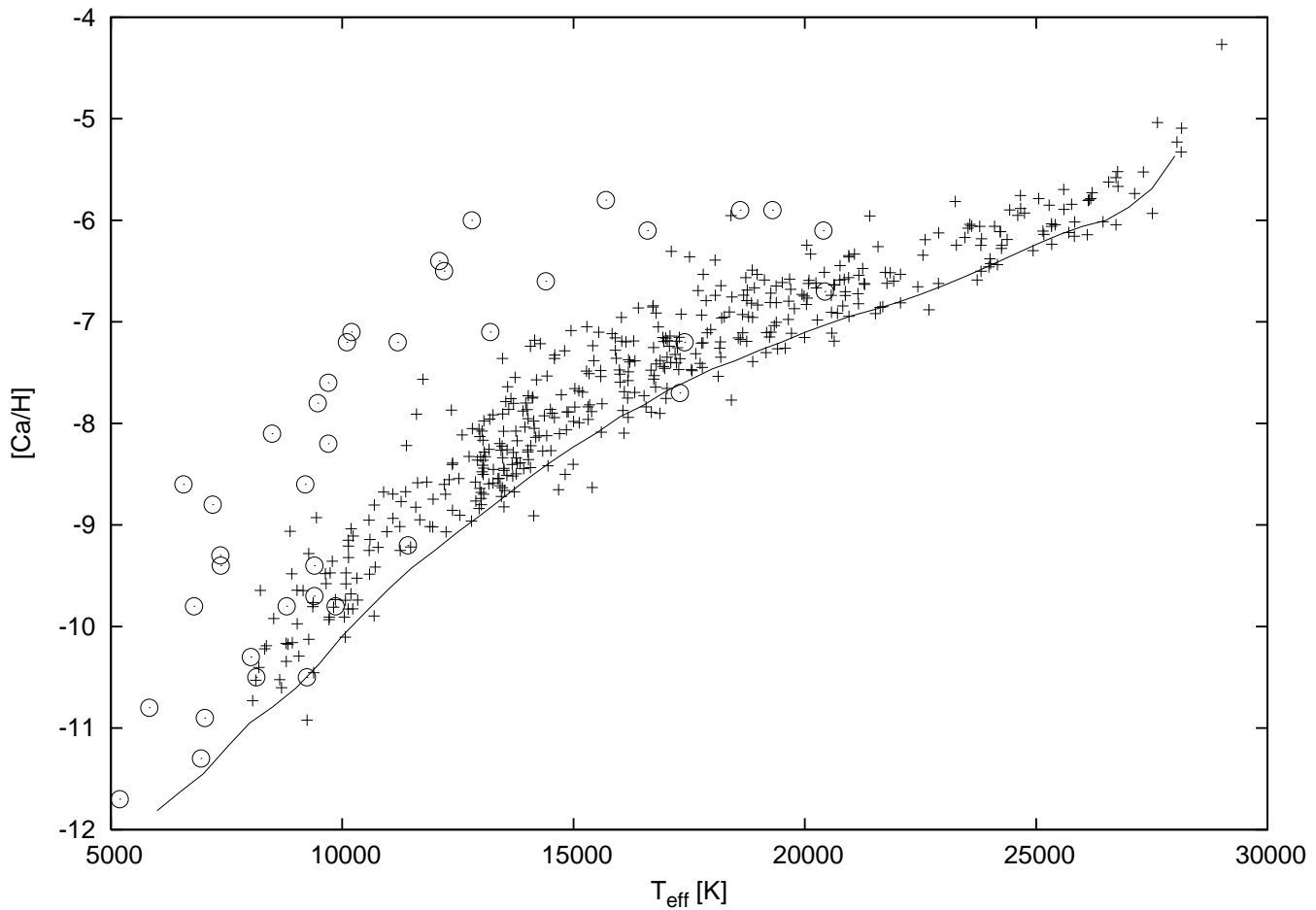}
\caption{Logarithmic Ca abundances (circles) and upper limits(crosses)
for the combined sample (see text). The continuous line indicates a
constant equivalent width of EW=15m\AA{}, taken as observational limit
for most objects \label{fig1}}
\end{figure*}

However, there are some exceptions to this rule. In cooler white
dwarfs below 25000~K, which are the topic of this study, we sometimes
find traces of Ca (through the strong CaII resonance lines),
occasionally accompanied by Mg and Fe. For He-rich atmospheres this
has been known from the time of the first white dwarfs discovered,
since the well known vMa~2 is a member of the class now called DZ (Z
always stands for the presence of metals, DZ means no other elements
visible). The class of hydrogen-rich objects with metals (DAZ, with
visible hydrogen and metal lines) has really only come into existence
over the past 8 years with the detection of a few dozen members in
large surveys at the Keck and VLT telescopes \citep{Zuckerman.Reid98,
Zuckerman.Koester.ea03, Koester.Rollenhagen.ea05}. The reason for this
difference is purely observational bias; because of the much lower
transparency of a H atmosphere at these temperatures, the CaII
equivalent widths in the DAZs are about a factor 1000 smaller than
in the DZs at the same abundance.

The diffusion timescales for the heavy elements to disappear from the
atmosphere -- or in case of an outer convection zone from the bottom
of this zone -- is always short compared to the cooling timescale. A
possibly competing process, radiative levitation, becomes completely
negligible for Ca below \Teff\ = 25000~K \citep{Chayer.Vennes.ea95}.
This implies that the observed metals cannot be primordial. They must
have been supplied from the outside, and currently the most widely
accepted mechanism for this is accretion of interstellar matter. There are
still some problems with this scenario, the most serious being the
absence of cool, dense clouds in the ISM of the solar
neighborhood. \cite{Aannestad.Kenyon.ea93} tried to correlate the
positions and motions of DZ (helium-rich) stars with conditions of the
ISM, but with inconclusive results. Alternative explanations are therefore
still being considered by several authors \citep[see e.g. the
  discussion in][]{Zuckerman.Koester.ea03}. 

Because the timescales for diffusion and thus metal visibility in the
DZs typically are on the order of 10$^6$~yr \citep[][ see also
Sect.~\ref{difftscales}]{Paquette.Pelletier.ea86*b}, the stars could
have traveled a large distance (of order 50~pc) since the accretion
ended, so that a lack of correlation is not too surprising. This situation
has changed dramatically with the recent detection of a large number
of DAZs. In these hydrogen-rich atmospheres the diffusion timescales
are about 3-4 orders of magnitude shorter, practically meaning that we
can only expect to observe metals if the accretion is still going on at
the present time. The DAZs thus offer two advantages compared to the
(helium-rich) DZs
\begin{itemize}
\item we know where the accretion occurs and can compare directly
  with the conditions of the ISM in that region
\item
we argue that we very likely observe a steady state
accretion/diffusion, meaning that we can determine the accretion rate
responsible for the observed abundances
\end{itemize}

There is thus a much more promising chance to learn
something about the accretion process and/or the conditions in the ISM
from an analysis of the observed metals in the DAZs.

\section{The accretion-diffusion scenario}
The idea that metals are accreted onto white dwarfs and then diffused
downward out of sight has been developed by many authors. It was
finally explored in great detail and transformed into a clear and
testable theory in a series of three fundamental papers by the
Montreal group
\citep{Dupuis.Fontaine.ea92,Dupuis.Fontaine.ea93,Dupuis.Fontaine.ea93*b}.
The basic idea of the so-called ``two-phase accretion model'' is that
the white dwarf travels for most of the time in the hot tenuous phase
of the ISM with negligible accretion rates. After typically $5 \times
10^7$ yr, it enters a dense cloud, where the accretion rate is high
and the metal abundance in the outer layers approaches an equilibrium
value between accretion and diffusion. After about $10^6$ yr
traveling through the cloud, the star emerges and the metal abundance
decreases exponentially with the diffusion timescale.

The observations available at that time -- almost exclusively of DZs
and DBZs -- agreed quite well with the predictions of this model,
although it obviously is an extreme oversimplification as emphasized
by the authors. It was subsequently called into question with the 
findings that the required dense neutral clouds, which would be
detectable through Na\,I absorption, are not found within the Local
Bubble, where these white dwarfs are located
\citep{Sfeir.Lallement.ea99, Welsh.Crifo.ea98, Welsh.Sfeir.ea99,
Redfield.Linsky02, Lehner.Jenkins.ea03}. Nevertheless we will keep
this model as our first hypothesis for this study since it is specific
enough to make quantitative, testable predictions.

While dense, cool gas seems to be absent in the Local Bubble, the
presence of warm, partially ionized gas has been clearly demonstrated
by \cite{Redfield.Linsky02, Redfield.Linsky04*b}. The
average number density of neutral hydrogen is around $n_{HI} = 0.1\
\mbox{cm}^{-3}$, but very little is known about the spatial
distribution. It is also unclear whether accretion under these
conditions would follow the Bondi-Hoyle rate
\citep[see][]{Koester.Rollenhagen.ea05}. However, following arguments
explained below, and in order to have a different model for testing
against the standard assumptions, we will define our alternative
hypothesis:
\begin{itemize}
\item the H-rich white dwarfs observed with metals are currently accreting
  in a steady state, with accretion  and diffusion proceeding at the
  same rate;
\item there are no large-scale differences (like the two phases of the
  standard model). Many, and perhaps all, white dwarfs in the Local
  Bubble are accreting with a continuous distribution of accretion
  rates caused by fluctuations in ISM densities and differences
  in the stellar parameters. The ones we can identify as DAZs are the top
  of the distribution.
\end{itemize}

Obviously this and the Dupuis et al. hypothesis are simplifications at
extreme ends of the possibilities and could be characterized as
``two-phase'' vs. ``continuous'' accretion. In the following sections
we explore whether the observations can lead to a distinction
between the alternatives.

\begin{figure}
\includegraphics[angle=270,width=0.48\textwidth]{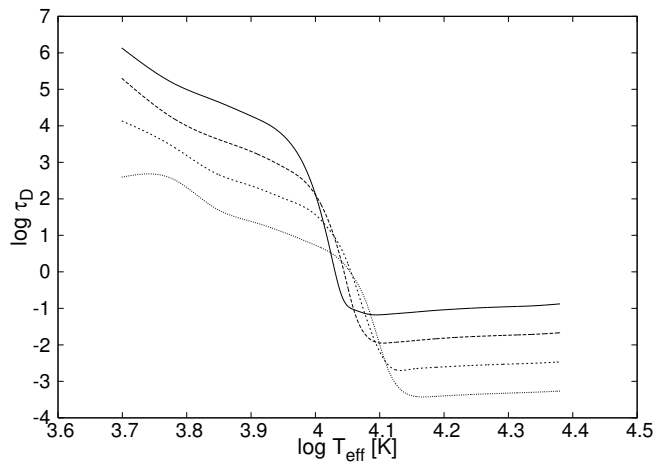}
\caption{Diffusion timescales (in years) of Ca. The solid line 
indicates \logg = 7.5, then following \logg = 8.0, 8.5, 9.0 \label{figtime}}
\end{figure}

\section{The observed sample}

The data come from two large surveys: a search for Ca in DAs using the
Keck telescope and the ESO Supernova Ia Progenitor Survey (SPY). The
instruments and reduction procedures are described in
\cite{Zuckerman.Koester.ea03} and
\cite{Koester.Rollenhagen.ea05}. Atmospheric parameters \Teff, \logg,
and Ca abundances are given in those papers and in
\cite{Berger.Koester.ea05}. Taking both sources together, we have a
sample of 38 DAZs  \citep[excluding those marked as double degenerates or
DA+DM pairs in][]{Zuckerman.Koester.ea03} with
observed photospheric CaII lines and about 450s DA with upper limits
for the Ca abundance.

Table~\ref{tab1} collects the data for the DAZs, while Fig.~\ref{fig1}
graphically shows the distribution of Ca abundances with \Teff. The line
corresponds to an equivalent width of the CaII~K line of 15~m\AA,
which is the observational limit for our spectra with the highest
signal-to-noise ratio. Because of the ionization of CaII to CaIII with
increasing effective temperature, the lower limit for observable Ca
increases until it becomes unobservable between 20000 and 25000~K. For
all objects with Ca lines in our sample, the photospheric nature (as
opposed to interstellar absorption) was confirmed through a comparison
of radial velocity determinations from the Ca and the hydrogen Balmer
lines.

Ca is found at all temperatures filling the area of abundances between
the lower observational limit and some upper limit [Ca/H]$ \approx -5.7$
([Ca/H] is the logarithmic number ratio of the elements). This upper limit
decreases towards lower \Teff, which will be discussed in a later
section.

\section{Diffusion timescales\label{difftscales}}
When trying to interpret the observational results in terms of an
accretion-diffusion scenario, the most important theoretical
ingredient is the diffusion timescale. In the absence of an outer
supply, heavy elements disappear exponentially out of the reservoir
(either the atmosphere or the outer convection zone, which is assumed
homogeneously mixed) on a timescale given by \cite[see
e.g.][]{Dupuis.Fontaine.ea92}
\begin{equation} 
 \tau_D = \frac{M_{WD}\, q}{4\pi r_b^2 \rho w} 
\end{equation}
where $M_{WD}$ is the mass of the white dwarf, $q$ the mass fraction
of the reservoir, $r_b$ the radius at the bottom of the reservoir
(always close to the white dwarf radius), $\rho$ the mass density, and
$w$ the relative diffusion velocity between the heavy trace
element and the main constituent, which is hydrogen in our case.   
\cite{Paquette.Pelletier.ea86*b} discuss the influence of the
parameters $q, \rho, w$ on the timescales, in particular with regard
to the large differences between He and H envelopes. Basically the
reason is the following: the convection zones ($q$) are much thicker
in He envelopes at the same mass and effective temperature. The
density at the bottom increases, but not as fast as $q$, and in
addition the velocity $w$ is smaller in denser environments. The
combination of these factors adds up to give timescales for He several
orders of magnitudes larger than for H at the same \Teff.

The Dupuis et al. papers only give timescales in helium atmospheres in
a tabular form, and only a few calculations for hydrogen envelopes below
10000~K, due to the lack of observational data at that
time. Although \cite{Paquette.Pelletier.ea86*b} give data for hydrogen
envelopes up to 15000~K, we nevertheless decided to repeat such
calculations in order to obtain a homogeneous set of data covering the
range of observations and incorporating the improved understanding of
the mixing-length parameters in DA white dwarfs.

For these calculations we used our stellar atmosphere and stellar
structure codes, which use up-to-date input physics for opacities and
non-ideal gas effects in the equation of state. The most relevant
fact for this study is the use of the mixing-length version ML2
instead of ML3, and a mixing length of 0.6 pressure scale
heights. Various mixing-length approximations used in white dwarf
modeling differ in the choice of three numerical constants of the
model \citep[see][ for a definition of the
nomenclature]{Fontaine.Villeneuve.ea81,Tassoul.Fontaine.ea90}. In
short, the ML3 version is a more efficient convection with higher
energy flux at the same temperature gradient.  Necessary stellar
parameters for the diffusion time calculations, e.g. the physical
conditions at the bottom of the reservoir of heavy elements to be
depleted, were obtained from these models.  Convection zones start
very shallow in DAs above \Teff $\approx$ 15000~K, and the reservoir
is the atmosphere. We define, somewhat arbitrarily, the limit at an
optical depth of $\tau_{Ross} = 5$. When a 0.6~\Msun\ white dwarf
cools down below 13000~K, the bottom becomes deeper than this, and the
reservoir is defined by the bottom of the convection zone. For a more
(less) massive white dwarf, the transition occurs at a slightly higher
(lower) \Teff.

Collision integrals for the diffusion velocities were obtained from
the fits in \cite{Paquette.Pelletier.ea86} and the calculations
followed the prescriptions in \cite{Paquette.Pelletier.ea86*b} very
closely. One major exception was that the detailed distribution over
ionization states was obtained from the models and average diffusion
velocities determined instead of just using the dominant ions as in
the latter work. In view of significant uncertainties in non-ideal gas
effects (e.g. pressure ionization) or convection theory, we do not
believe that the result will be more accurate, but this method avoids
discontinuities in timescale changes with temperature of the
models. Table~\ref{tabtime} gives the resulting timescales for Ca, Mg,
and Fe in hydrogen, while Fig.~\ref{figtime} shows the Ca data in graphic
form.  In general these results are very similar to those obtained by
\cite{Paquette.Pelletier.ea86*b} and earlier work
\citep{Muchmore79,Muchmore84}. The main difference is that the steep
increase in diffusion timescales towards cooler temperatures occurs at
slightly lower \Teff\ because of different formulations of the
mixing-length theory used, e.g. ML3 in the case of
\cite{Paquette.Pelletier.ea86*b}.

The first conclusion from these data is a confirmation of earlier
results: the diffusion timescales are extremely short compared to
evolutionary timescales, and also much shorter than diffusion timescales
in He-rich envelopes. Over most of the observed range, they are much
shorter than 1000~yr, going down to days at the high-temperature end;
and only for the very coolest objects do the timescales approach
100000~yr. The steep increase in the timescales between 12500 and
10000~K is directly connected to the very strong increase in the depth
of the convection zone, when the white dwarf cools through this
temperature region.
If the distribution in Fig.~\ref{fig1} is due to objects
{\em after} the end of their accretion episode (note that an exponential
decline of a metal abundance means equal time for each decade of
abundances and thus a ``homogeneous'' population of the area above the
visibility limit), then the metals will be visible only for a very
short time compared to the cooling time through this range -- approximately
$5.6 \times 10^8$ yr from 20000 to 10000~K, and $1.7 \times 10^9$ yr from 10000
to 6000~K \citep{Wood95}. This conflicts with the result that a
significant fraction of DA shows metals, $>20\%$ in
\cite{Zuckerman.Koester.ea03} and still $>5\%$ in
\cite{Koester.Rollenhagen.ea05}, a sample much more biased towards
higher \Teff. This high fraction is only possible with a constant
supply of new DAZs through accretion. On the other hand, taking the
numbers for the two-phase scenario at face value, where the star
spends on the order of $10^6$ yr in the accretion phase, one would
expect a completely different distribution with the vast majority of
the objects near the upper boundary of abundances.  This leaves little
doubt that this interpretation is not correct and that we have to
assume that accretion is still ongoing in most, if not all
objects. This conclusion is inevitable for the hot objects with
timescales of a few days (e.g. about 10 days for G29-38, which has
been observed to have Ca since 1997), but is very plausible for all
objects. We will thus assume that all DAZ white dwarfs are currently
accreting at a steady state (between accretion and diffusion), with
this generalization made plausible by the short diffusion
timescales \citep[see][ for a discussion]{Dupuis.Fontaine.ea92}.

\begin{table*}[tbp]
\caption{Logarithm of the diffusion timescales (in years) for Ca, Mg, and Fe in
  hydrogen-rich white dwarfs as a function of  effective temperature and
  surface gravity \label{tabtime}}
\begin{small}
\begin{center}
\begin{tt}
\begin{tabular}{rr|rrrrrrr}
          &  &\multicolumn{7}{c}{ \logg}   \\ 
 \Teff  [K] && 7.5       &  7.75      &  8.0       &  8.25      &  8.5       &  8.75      &  9.0     \\ \hline  
            &Ca& 6.1353  &  5.7436    &   5.3045    &  4.7835   &  4.1353   &   3.4683   &  2.5939    \\
5000        &Mg& 6.1629  &  5.7188    &   5.2500    &  4.8033   &  4.2051   &   3.5215   &  2.6560    \\
            &Fe& 6.1406  &  5.7159    &   5.2216    &  4.6712   &  3.9877   &   3.3064   &  2.4714    \\ 
            &Ca& 5.1778  &  4.6687    &   4.2179    &  3.8380   &  3.4399   &   3.0440   &  2.5471    \\
6000        &Mg& 5.2526  &  4.7300    &   4.2702    &  3.8767   &  3.4632   &   3.0424   &  2.5036    \\
            &Fe& 5.1932  &  4.6841    &   4.2439    &  3.8780   &  3.4981   &   3.1134   &  2.5948    \\ 
            &Ca& 4.6835  &  4.1794    &   3.6554    &  3.1968   &  2.7003   &   2.2280   &  1.7375    \\
7000        &Mg& 4.7280  &  4.2117    &   3.6735    &  3.2086   &  2.6994   &   2.2151   &  1.7150    \\
            &Fe& 4.6304  &  4.1124    &   3.5628    &  3.0924   &  2.5650   &   2.0642   &  1.5538    \\ 
            &Ca& 4.2452  &  3.7668    &   3.2781    &  2.8298   &  2.3413   &   1.8960   &  1.3673    \\
8000        &Mg& 4.2724  &  3.7838    &   3.2839    &  2.8307   &  2.3366   &   1.8874   &  1.3521    \\
            &Fe& 4.1262  &  3.6333    &   3.1189    &  2.6570   &  2.1427   &   1.6812   &  1.1178    \\ 
            &Ca& 3.6618  &  3.2656    &   2.8340    &  2.4193   &  1.9883   &   1.5547   &  1.0580    \\
9000        &Mg& 3.6729  &  3.2722    &   2.8328    &  2.4146   &  1.9811   &   1.5451   &  1.0462    \\
            &Fe& 3.4729  &  3.0716    &   2.6270    &  2.2022   &  1.7571   &   1.3075   &  0.7892    \\ 
            &Ca& 2.1235  &  2.1596    &   2.1125    &  1.8695   &  1.5642   &   1.1934   &  0.7282    \\
10000       &Mg& 2.0589  &  2.1135    &   2.0940    &  1.8610   &  1.5580   &   1.1841   &  0.7134    \\
            &Fe& 1.9373  &  1.9567    &   1.9125    &  1.6605   &  1.3344   &   0.9397   &  0.4499    \\ 
            &Ca& 0.7272  &  1.0429    &   1.3574    &  1.3264   &  1.1629   &   0.9366   &  0.5481    \\
10500       &Mg& 0.7107  &  0.9840    &   1.2991    &  1.2851   &  1.1402   &   0.9232   &  0.5323    \\
            &Fe& 0.6606  &  0.9191    &   1.1606    &  1.1139   &  0.9485   &   0.7065   &  0.2875    \\ 
            &Ca&-0.6754  & -0.2462    &   0.2184    &  0.4745   &  0.5944   &   0.5090   &  0.2723    \\
11000       &Mg&-0.4665  & -0.1343    &   0.1712    &  0.4423   &  0.5441   &   0.4723   &  0.2487    \\
            &Fe&-0.8266  & -0.2525    &   0.1212    &  0.3335   &  0.3868   &   0.2845   &  0.0385    \\ 
            &Ca&-1.0396  & -1.2166    &  -1.0994    & -0.5974   & -0.2729   &  -0.1556   & -0.1807    \\
11500       &Mg&-0.8548  & -1.0199    &  -0.8862    & -0.5751   & -0.3171   &  -0.1855   & -0.2240    \\
            &Fe&-1.2632  & -1.3974    &  -1.1010    & -0.6656   & -0.4005   &  -0.3263   & -0.4012    \\ 
            &Ca&-1.1486  & -1.4876    &  -1.7497    & -1.6973   & -1.2673   &  -0.9217   & -0.8417    \\
12000       &Mg&-0.9675  & -1.3054    &  -1.5580    & -1.4801   & -1.1788   &  -0.9544   & -0.8612    \\
            &Fe&-1.3827  & -1.7171    &  -1.9453    & -1.7734   & -1.3084   &  -1.0355   & -0.9847    \\ 
            &Ca&-1.1597  & -1.5499    &  -1.9446    & -2.2830   & -2.5642   &  -2.8837   & -2.7075    \\
13000       &Mg&-0.9673  & -1.3627    &  -1.7644    & -2.0971   & -2.3796   &  -2.6888   & -2.5069    \\
            &Fe&-1.3675  & -1.7691    &  -2.1787    & -2.5123   & -2.7764   &  -3.0656   & -2.7119    \\ 
            &Ca&-1.1156  & -1.5065    &  -1.8966    & -2.2858   & -2.6730   &  -3.0616   & -3.3721    \\
14000       &Mg&-0.9063  & -1.3008    &  -1.6961    & -2.0920   & -2.4880   &  -2.8880   & -3.1820    \\
            &Fe&-1.2870  & -1.6861    &  -2.0857    & -2.4867   & -2.8887   &  -3.2975   & -3.5648    \\ 
            &Ca&-1.0700  & -1.4602    &  -1.8491    & -2.2354   & -2.6268   &  -3.0241   & -3.4227    \\
15000       &Mg&-0.8583  & -1.2519    &  -1.6451    & -2.0368   & -2.4286   &  -2.8253   & -3.2286    \\
            &Fe&-1.2172  & -1.6147    &  -2.0126    & -2.4102   & -2.8097   &  -3.2097   & -3.6109    \\ 
            &Ca&-1.0351  & -1.4223    &  -1.8116    & -2.2034   & -2.5981   &  -2.9955   & -3.3929    \\
16000       &Mg&-0.8205  & -1.2101    &  -1.6031    & -1.9952   & -2.3871   &  -2.7870   & -3.1872    \\
            &Fe&-1.1619  & -1.5572    &  -1.9569    & -2.3555   & -2.7510   &  -3.1483   & -3.5439    \\ 
            &Ca&-1.0070  & -1.3937    &  -1.7838    & -2.1772   & -2.5744   &  -2.9721   & -3.3693    \\
17000       &Mg&-0.7924  & -1.1822    &  -1.5741    & -1.9612   & -2.3587   &  -2.7580   & -3.1572    \\
            &Fe&-1.1056  & -1.5030    &  -1.9037    & -2.2994   & -2.6988   &  -3.0968   & -3.4931    \\ 
            &Ca&-0.9846  & -1.3722    &  -1.7646    & -2.1596   & -2.5554   &  -2.9530   & -3.3503    \\
18000       &Mg&-0.7697  & -1.1600    &  -1.5492    & -1.9401   & -2.3369   &  -2.7358   & -3.1344    \\
            &Fe&-1.0435  & -1.4465    &  -1.8485    & -2.2492   & -2.6490   &  -3.0523   & -3.4543    \\ 
            &Ca&-0.9681  & -1.3565    &  -1.7488    & -2.1438   & -2.5397   &  -2.9373   & -3.3346    \\
19000       &Mg&-0.7512  & -1.1406    &  -1.5288    & -1.9234   & -2.3199   &  -2.7184   & -3.1167    \\
            &Fe&-0.9809  & -1.3866    &  -1.7916    & -2.1955   & -2.6010   &  -3.0076   & -3.4125    \\ 
            &Ca&-0.9545  & -1.3433    &  -1.7352    & -2.1304   & -2.5263   &  -2.9239   & -3.3213    \\
20000       &Mg&-0.7364  & -1.1252    &  -1.5148    & -1.9102   & -2.3065   &  -2.7047   & -3.1027    \\
            &Fe&-0.9241  & -1.3308    &  -1.7373    & -2.1447   & -2.5525   &  -2.9617   & -3.3692    \\ 
            &Ca&-0.8771  & -1.2732    &  -1.6684    & -2.0677   & -2.4669   &  -2.8662   & -3.2657    \\
24000       &Mg&-0.7030  & -1.0903    &  -1.4830    & -1.8772   & -2.2719   &  -2.6675   & -3.0640    \\
            &Fe&-0.7958  & -1.1973    &  -1.6004    & -2.0050   & -2.4111   &  -2.8187   & -3.2277    \\ 
\end{tabular}
\end{tt}
\end{center}
\end{small}
\end{table*}

\section{Steady state accretion rates}
\begin{figure}
\includegraphics[angle=270,width=0.48\textwidth]{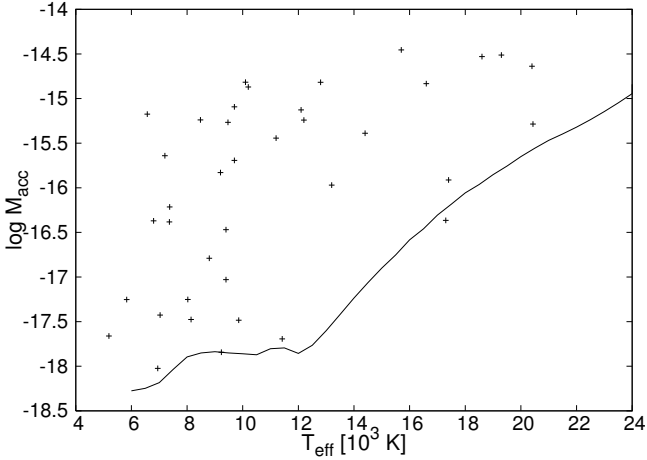}
\caption{Accretion rates for 38 DAZ in solar masses per year, assuming
  that the Ca abundance in the accreted material is solar. The line shows
the limit of the accretion rates resulting from an observational limit
of 15m\AA{} for a 0.6~\Msun\ white dwarf
\label{figaccretion}} 
\end{figure}

With the assumption of steady state accretion, we can immediately
determine the accretion rate \citep{Dupuis.Fontaine.ea93}
\begin{equation}  
          \dot{M}_{acc} = \frac{M_{WD}\, q X}{\tau_{D} X_{acc}}
\end{equation}
with mass fraction of the heavy element in the reservoir $X$ and its
abundance in the accreting matter $X_{acc}$. We assumed the solar
abundance of Ca for $X_{acc}$ and took the mass fraction $q$ of the
reservoir from our models with the appropriate stellar parameters. The
resulting mass accretion rates are given in Table~\ref{tabaccretion}
and Fig.~\ref{figaccretion}.

\begin{table}[ht]
\caption{Diffusion timescales $\tau_D $,
mass fraction at the layer of diffusion, $\log q  $, and the accretion
rates $\dot{M}$ in \Msun/yr for 38 DAZs
abundances.\label{tabaccretion}} 
\begin{center}
{\tt
\begin{tabular}{rrrrr}
 Object      & $\log\tau_D$ & $M_{WD}$ &  $\log q$   & $\log\dot{M}$
 \\ 
\hline
 WD0032-175  &    2.72  &   0.600   &   -10.04 &   -17.84   \\
 HS0047+1903 &   -1.48  &   0.512   &   -15.56 &   -14.83   \\
 HE0106-3253 &   -1.82  &   0.616   &   -15.90 &   -14.45   \\
 WD0208+296  &    3.78  &   0.537   &   -8.43  &   -15.64   \\
 WD0235+064  &   -1.04  &   0.550   &   -14.91 &   -17.69   \\
 WD0243-026  &    3.39  &   0.719   &   -8.67  &   -16.37   \\
 WD0245+541  &    4.47  &   0.712   &   -6.98  &   -17.66   \\
 HS0307+0746 &    1.81  &   0.664   &   -11.42 &   -14.87   \\
 WD0408-041  &   -1.56  &   0.506   &   -15.70 &   -15.39   \\
 WD0543+579  &    3.22  &   0.597   &   -9.17  &   -17.48   \\
 WD0846+346  &    3.51  &   0.595   &   -8.72  &   -16.22   \\
 WD1015+161  &   -1.59  &   0.571   &   -15.60 &   -14.51   \\
 WD1102-183  &    3.27  &   0.597   &   -9.10  &   -17.25   \\
 WD1116+026  &   -1.73  &   0.552   &   -15.86 &   -15.24   \\
 WD1124-293  &    2.29  &   0.663   &   -10.66 &   -15.69   \\
 WD1150-153  &   -1.63  &   0.500   &   -15.79 &   -14.82   \\
 WD1202-232  &    2.59  &   0.724   &   -9.900 &   -16.79   \\
 WD1204-136  &   -0.34  &   0.606   &   -14.01 &   -15.44   \\
 WD1208+576  &    4.50  &   0.530   &   -7.31  &   -17.25   \\
 HE1225+0038 &    2.48  &   0.662   &   -10.31 &   -17.03   \\
 WD1257+278  &    3.26  &   0.541   &   -9.25  &   -15.24   \\
 HE1315-1105 &    2.02  &   0.857   &   -10.63 &   -16.47   \\
 WD1337+705  &   -1.57  &   0.574   &   -15.56 &   -15.29   \\
 WD1344+106  &    3.68  &   0.594   &   -8.46  &   -18.02   \\
 WD1407+425  &    2.27  &   0.602   &   -10.83 &   -17.48   \\
 WD1455+298  &    4.33  &   0.386   &   -7.98  &   -16.38   \\
 WD1457-086  &   -1.73  &   0.628   &   -15.71 &   -14.64   \\
 WD1614+160  &   -1.46  &   0.514   &   -15.53 &   -15.91   \\
 WD1633+433  &    3.69  &   0.654   &   -8.34  &   -15.17   \\
 WD1821-131  &    2.89  &   0.853   &   -9.21  &   -17.43   \\
 WD1826-045  &    2.58  &   0.661   &   -10.11 &   -15.83   \\
 WD1858+393  &    2.58  &   0.601   &   -10.30 &   -15.27   \\
 WD2115-560  &    2.29  &   0.663   &   -10.66 &   -15.09   \\
 HS2132+0941 &   -1.47  &   0.454   &   -15.63 &   -15.97   \\
 WD2149+021  &   -1.62  &   0.565   &   -15.68 &   -16.37   \\
 HE2221-1630 &    1.84  &   0.727   &   -11.28 &   -14.82   \\
 HS2229+2335 &   -1.60  &   0.569   &   -15.62 &   -14.53   \\
 WD2326+049  &   -1.70  &   0.551   &   -15.81 &   -15.13   \\ \hline
\end{tabular}
}
\end{center}
\end{table}

The continuous distribution of observed Ca abundances thus translates
into a continuum of derived accretion rates, with a maximum rate of
approximately $3 \times 10^{-15}$ \Msun/yr over the whole range of observed
temperatures. The lack of higher abundances below \Teff\ $\approx$
10000~K in Fig.~\ref{fig1} is due to the bottom of the reservoir moving
into deeper and denser regions, as can be seen from the combination of
Eqs.(1) and (2):
\begin{equation}
          X = \frac{\dot{M}_{acc} X_{acc} \tau_D}{M_{WD}\,q}
            = \frac{\dot{M}_{acc} X_{acc}}{4\pi r_b^2 \rho w}.
\end{equation}
The radius does not change significantly, but the increased diffusion
flux $\rho w$ leads to lower abundances in the observable reservoir.
The range of accretion rates is in good agreement with the results of
\cite{Dupuis.Fontaine.ea93*b} for helium-envelope white dwarfs; in
fact, their ``high accretion rate'' used for the comparison with
observed abundances is $5 \times 10^{-15}$~\Msun/yr, in excellent agreement
with Fig.~\ref{figaccretion}. If the maximum accretion rates are
indeed independent of the surface temperature of the star, one could
even speculate that our mixing-length description predicts convection
zone depths at the low temperature end, which are slightly too large!

\section{Correlation of accretion rates with stellar parameters}
 What is the reason for the large variation in accretion rates,
  e.g. by three orders of magnitude at the cool end? To study this
  question, we have to proceed to the theoretically much less secure
  ground of the accretion process. Rather than discuss the problems
  of the physics of accretion, which is beyond the scope of the
  present study, we assume that the mass accretion rate is given
  by the Bondi-Hoyle formulation
  \citep{Hoyle.Lyttleton39,Bondi.Hoyle44, Bondi52, Ruffert94,
  Ruffert.Arnett94}:
\begin{equation}
    \dot{M}_{acc} = \frac{4\pi G^2 M^2 \rho}{(c^2 +
    v_s^2)^{3/2}}. 
\end{equation} 
Here, $M$ is the stellar mass, $v_s$ its velocity relative to
the interstellar gas at large distance, $c$ the sound velocity
and $\rho$ the unperturbed ISM density. For the warm ($\approx$
8000~K) phase of the ISM, the sound velocity is not important compared
to the space velocity for most objects; we have taken $c = 8$ km/s 
and use a new variable $v = \sqrt{c^2+v_s^2}$ in the following. The accretion
rate therefore can be separated into individual properties of the
stars ($\propto M^2/v^3$) and the condition of the ambient medium
given by $\rho$. Fortunately, the stellar parameters mass and space
velocity (in 8 objects estimated from the radial velocity) are known
for our complete sample. 

\begin{figure}
\includegraphics[angle=270,width=0.48\textwidth]{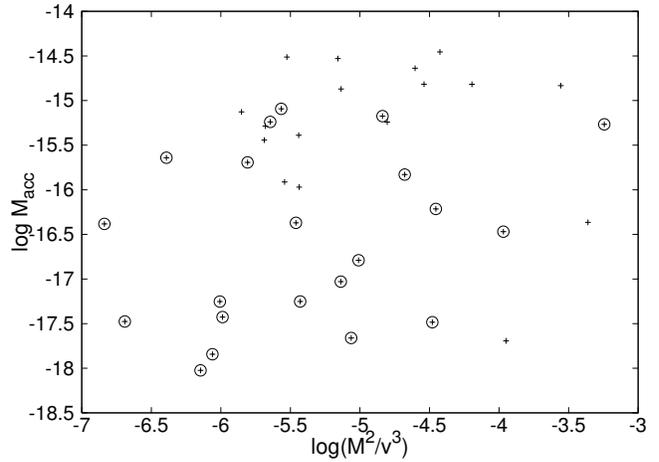}
\caption{Accretion rates vs. the ``stellar factor'' $M^2/v^3$ for 38
objects. Objects with \Teff $<$ 10000 K are marked with
circles. Accretion rates are in \Msun\ per year, the units for the
stellar factor are \Msun\ and km/s.
\label{figstellpar}} 
\end{figure}

\begin{figure}
\includegraphics[angle=270,width=0.48\textwidth]{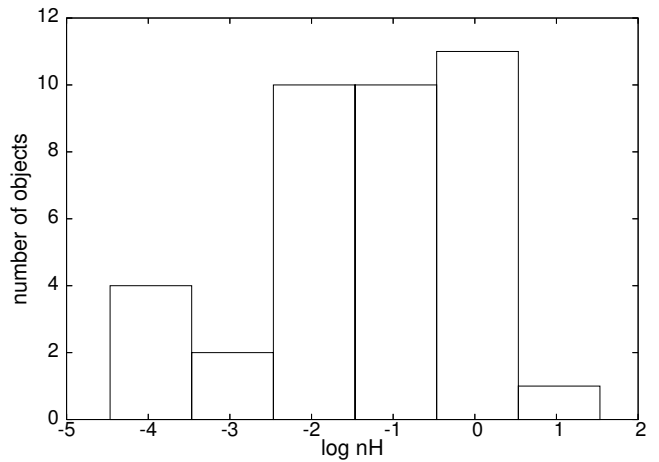}
\caption{ISM density (neutral + ionized hydrogen per cm$^3$) derived from
  assuming the validity of the Bondi-Hoyle accretion formula. See text
\label{figdens}} 
\end{figure}

Figure~\ref{figstellpar} shows accretion rates as a function of the
combination $M^2/v^3$ (``stellar factor''), which enters the
Bondi-Hoyle formula Eq. (4). Although the stellar factor varies by
more than three orders of magnitude, there is obviously no correlation
with the accretion rates. At most one could state that there is a
small tendency towards higher rates for large factors and vice versa.
This can only mean that the second factor, the interstellar density,
must play an equally large role and show similar variations if
accretion is coming from the ISM. This is demonstrated in
Fig.~\ref{figdens}, which shows the distribution of interstellar
density (derived from all the other known factors in Eq. (4) and
converted to hydrogen particle density) as a histogram. As expected
from the lack of correlation in Fig.~\ref{figaccretion}, the derived
interstellar densities span the range $n_H = 0.01 - 1$ cm$^{-3}$, with
very few objects outside (we use $n_H$ for the sum of neutral and
ionized hydrogen atoms). It should be noted that this number is
derived from the mass density and thus includes neutral and ionized
hydrogen. The direct (not log) average of all values is $\approx 0.5$,
which is very likely biased towards higher values, because for higher
\Teff\ we can observe only the highest accretion rates.

For most of the current sample the distances are known, and we can
therefore determine the location in the solar neighborhood
corresponding to the individual density determinations. We can also
derive upper limits for the ISM densities for many of the numerous
objects with upper limits for the Ca abundance (Fig.~\ref{fig1}). We
had hoped to find examples of several DAZs close together in an area
of the sky, all with enhanced ISM densities derived from the accretion
rates, but no such correlations are apparent. If our interpretation up
to this point is at all meaningful, this means that the ISM density
varies by two orders of magnitude on the scale of a few pc.

\section{Discussion and conclusions}
The two-phase accretion scenario in the highly idealized form of the Dupuis
et al. work cannot explain the observations of the DAZs discussed in
this study. The absence of cold neutral matter in the solar
neighborhood, the large fraction of DA showing metals in spite of short
diffusion timescales, and the distribution of the observed abundances
all point to ongoing accretion even under the conditions of the ISM in
the Local Bubble. 

Although we show that the stellar parameters entering the calculation
of accretion rates according to Bondi-Hoyle hydrodynamic accretion
vary over several orders of magnitude, this alone cannot explain the
variation in observed accretion rates. Additionally, we need a
significant variation in the ISM densities, with typical values of
ionized plus neutral hydrogen
$n_H \approx 0.1 - 0.2$~cm$^{-3}$ and variation by a factor of 10 in both
directions. The only viable candidate is the warm, partially ionized
phase of the solar neighborhood ISM, which has been studied recently
in a series of papers by \cite{Linsky.Redfield.ea00} and
\cite{Redfield.Linsky00, Redfield.Linsky02, Redfield.Linsky04*b,
Redfield.Linsky04}.

From these papers the following picture emerges. The absorption lines
along the line-of-sight to many nearby stars show the existence of
distinct components, tentatively called ``clouds''. Neutral hydrogen
densities are typically $n_{HI} \approx 0.1$ cm$^{-3}$, with hydrogen
ions adding $n_p \approx 0.11$ cm$^{-3}$ to the mass. The length scale
is 2.2 pc, with a range of 0.1 to 11 pc. Our own sun lies within a
similar cloud, the Local Interstellar Cloud or LIC, with extensions of
4 - 6 pc. We use these numbers for some very simple estimates, to test
the plausibility of a connection between these clouds and the DAZ
phenomenon.

Assuming current accretion within such a cloud to be responsible for
the observed metals, the observations demand a filling factor on the
order of 10\% within 50~pc. Taking spherical clouds with a radius of
1.5~pc, the total number of clouds in this volume would be 3700. The
total solid angle covered by these clouds (distributed homogeneously)
is 31.4 or 2.5 times the total sphere of $4\pi$. This is in excellent
agreement with the findings of 1-3 components (and an average of
$\approx 2$) along each line-of-sight in the Redfield and Linsky
studies.  A white dwarf with a space velocity of 30~km/s would travel
1~pc in about 33000~yr, plenty of time to reach steady state
abundances and stay constant over the observational timescales.

The typical distance between the centers of any two clouds would be
3.2~pc, about the same as the size of the clouds. This should serve as
a caveat and a reminder that the concept of ``clouds'' may again be an
oversimplification. However, this does not affect our interpretation,
since all we need is a scale length for significant variation in the
physical parameters of the ISM. These estimates are all quite
favorable to the hypothesis of ongoing accretion from the warm phase
of the ISM, alleviating the strongest objection against the
accretion/diffusion scenario -- the absence of cool, neutral clouds.

There are, however, other open questions remaining. One of them is
whether the assumption of fluid (Bondi-Hoyle) accretion rates is
justified. \cite{Koester76} concluded that accretion rates would be
much lower than the fluid rate since interactions are not sufficient
to destroy the momentum perpendicular to the accretion column, a
necessary condition for the hydrodynamic treatment. On the other hand,
\cite{Alcock.Illarionov80} argued that an ionized plasma would always
accrete at the fluid rate due to plasma instabilities and magnetic
fields.

Another difficult question is the composition of the accreted
matter. We have derived a very robust result on the accretion of Ca
atoms (which can be derived from the mass accretion rates in
Table~\ref{tabaccretion} by multiplying those numbers by
$8 \times 10^{-5}$). The total accretion rate follows from {\em
assuming} a solar composition. There is no evidence for that; on the
contrary, observations in helium-rich white dwarfs with metals
indicate that in many cases very little or no hydrogen is
accreted. This seems to show that the accretion is preferentially of
dust grains, either interstellar or from some circumstellar cloud or
disk. Circumstellar material has been found for G29-38 and GD362
\citep[see e.g.][ for recent studies]
{Jura03,Reach.Kuchner.ea05,Becklin.Farihi.ea05,Kilic.von-Hippel.ea05},
with the origin attributed to comets, tidal disruption of planets or
asteroids, or the merging of two white dwarfs
\citep{Livio.Pringle.ea05}. It is quite possible that in such cases
accretion from a different source than the ISM is responsible for the
observed metals; similarly, in several cases of close M dwarf
companions \citep{Zuckerman.Koester.ea03}, a stellar wind may provide
the heavy elements. 

However, these seem to be exceptional cases. As found in the
comprehensive recent work by \cite{Farihi.Becklin.ea05} none of the
371 white dwarfs studied showed a near-infrared excess indicative of
circumstellar dust similar to G29-38 and GD362. One should notice that
the search was limited by the signal-to-noise ratio of the Ks
magnitude in the 2MASS sample and only about 1/3 of the sample had
adequate data, so further detections may be forthcoming from ongoing
searches (J. Farihi, priv. comm.).  Nevertheless, it appears unlikely
that such processes can account for the 10 - 20\% of hydrogen-rich
white dwarfs that are below 20000~K and have heavy elements.

\acknowledgement{We are grateful to Ben Zuckerman and Michael Jura for
  stimulating discussions. This work was supported in part by a grant
  from the Deutsche Forschungsgemeinschaft (Ko738/21-1,2).}

%\bibliographystyle{aa}
%\bibliography{wdnorm_1,wdnorm_2,wdnorm_3}

\end{document}